\documentclass[letterpaper,twocolumn,10pt]{article}
\usepackage{usenix}

\usepackage{verbatim}
\usepackage{amstext}
\usepackage{graphicx}
\usepackage{amssymb}
\usepackage{balance}

\makeatletter

\newcommand{\supsym}[1]{\raisebox{4pt}{{\footnotesize #1}}}
\newcommand{\gt}{\supsym{$\dag$}}
\newcommand{\tf}{\supsym{$\ddag$}}
\newcommand{\cm}{\supsym{$\S$}}

\usepackage{url}

\@ifundefined{showcaptionsetup}{}{%
 \PassOptionsToPackage{caption=false}{subfig}}
\usepackage{subfig}
\makeatother

\begin{document}
\title{\Large \bf Can User-Level Probing Detect and Diagnose\\Common Home-WLAN Pathologies?}

\author{
{\rm Partha Kanuparthy~\gt\thanks{Contact author: {\tt partha@cc.gatech.edu}}, Constantine Dovrolis~\gt, Konstantina Papagiannaki~\tf,}
\and {\rm Srinivasan Seshan~\cm, Peter Steenkiste~\cm}\\
\begin{small}
{\gt~Georgia Institute of Technology~~~~ \tf~Telefonica Research~~~~ \cm~Carnegie Mellon University}
\end{small}
}

\maketitle
\begin{abstract}
Common Wireless LAN (WLAN) pathologies include low signal-to-noise ratio, congestion,
hidden terminals or interference from non-802.11 devices and phenomena.
Prior work has focused on the detection and diagnosis of such problems
using layer-2 information from 802.11 devices and special-purpose
access points and monitors, which may not be generally available.
Here, we investigate a user-level approach: is it possible to detect
and diagnose 802.11 pathologies with strictly user-level active probing,
without any cooperation from, and without any visibility in, layer-2
devices? In this paper, we present preliminary but promising
results indicating that such diagnostics are feasible.
\end{abstract}

\section{Introduction}

Most home networks today use an 802.11 Wireless LAN (WLAN) 
with a single Access Point
(AP), typically operating in Distributed Coordination Function (DCF)
mode. Home WLANs often suffer from various performance pathologies,
such as low signal strength, significant noise, interference from
external non-802.11 devices and physical phenomena, various forms
of fading, hidden terminals from devices in the same WLAN or in nearby
WLANs, or congestion. These pathologies can result in throughput degradation,
significant jitter and packet losses. To make things worse, due to the wireless
nature of the medium, troubleshooting WLAN performance is hard even
for experts, leave alone home users.

User-level probing is a well-established research area in wired networks
and it is used in practice to infer various properties and 
problems in such networks.
In the wireless domain, on the other hand, it is still unclear whether 
user-level probing can be nearly as effective.
A main motivation behind this work is to answer the following
``intellectual curiosity'' question: 
{\em is it possible to diagnose common WLAN performance problems
using active probing, without any information from, or modifications
to, the 802.11 devices or AP?} 
The methods presented in this paper show promising (but preliminary) results, 
potentially opening a new research thread within the area of wireless networks.

Specifically, our objective is to construct a
user-level tool for any 802.11 DCF WLAN that can detect: 
a) low Signal-to-Noise Ratio (SNR),
b) Hidden Terminals (HT), or c) congestion.
The methodology we propose, referred to as {\em WLAN-probe},
is a simple, easy-to-use, client-server application that eliminates
the need for vendor-specific network card (NIC), driver, AP, 
monitoring devices, or network
modifications. It is also portable across platforms, since it only requires
a user-level socket library (e.g., Berkeley sockets, POSIX, or Winsock APIs).

There are several reasons for a user-level probing tool:
\begin{description}
\item [Usability:] We want to build a diagnostic tool that would not 
require the user to install a specific NIC, AP, or modify the kernel 
(moreover, it would not require administrative privileges).
The user would just run a \emph{single instance} of the WLAN-probe client 
at the wireless link that appears problematic.
\item {\bf Hardware-agnostic:} Most wireless cards today export some form
of signal strength; for example, the Received Signal Strength Indicator (RSSI).
RSSI implementations are vendor-specific and they are not uniform across NICs. 
A user-level approach avoids the need to calibrate NIC statistics 
across devices and drivers on different OSes.
\item {\bf Software-agnostic:} A user-level approach eliminates the need to
write and maintain a hardware-compatibility layer for different OSes that
would expose NIC statistics at user-space. An example of that approach is 
WRAPI \cite{WRAPI}, designed to work on Windows XP with NICs supporting 
NDIS 5.1 drivers.
\item {\bf Passive inference:} Understanding active probing in the wireless 
domain may also enable methods for passive inference.
For example, is it possible to troubleshoot client performance at a remote 
web or video server using strictly application traffic? 
\end{description}
State-of-the-art diagnosis tools require (or modify) 
vendor-specific drivers and NICs, or they require special monitors at
the home network; we cover these approaches in the related work section.

WLAN-probe is based on two fundamental effects: 
a) the fact that low-SNR and HTs cause a dependency between
packet size and retransmission probability, while congestion does
not do so, and b) the fact that low-SNR conditions differ significantly
from HTs in the delay or loss temporal correlations they create. 
However, measuring layer-2 retransmissions and delays is not feasible
without information from the link layer. In
this short paper, we present the basic ideas and algorithms for 
user-level inference of link layer effects, with a
limited testbed evaluation. In future work,  we will conduct 
a more extensive evaluation, experiment with actual deployment at several
home networks, and expand the set of diagnosed pathologies.

\begin{figure}
\begin{centering}
\includegraphics[width=0.4\textwidth]{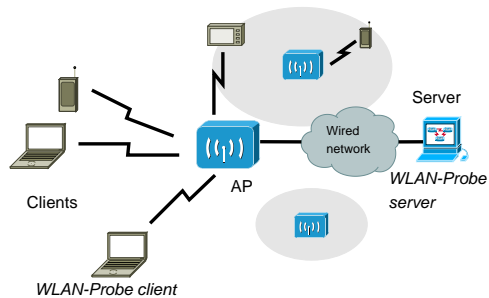}
\par\end{centering}
\caption{System architecture.\label{fig:System-architecture.}}
\end{figure}

We consider the following architecture, which is typical for most
home WLANs (see Figure \ref{fig:System-architecture.}). A single
802.11 AP is used to interconnect a number of wireless
devices; we do not make any assumptions about the exact type of the
802.11 devices or AP.
We assume that another computer, used as our WLAN-probe measurement
server $S$ is connected to the AP through an Ethernet connection.
This is not difficult in practice given that most APs provide an Ethernet
port, as long as the user has at least two computers at home. 
The key requirement for the server $S$ and its connection to the WLAN AP
is that it should not introduce significant jitter (say more than 1-3msec).
The server $S$ allows us to probe the WLAN channel without demanding 
\emph{ping-like} replies from the AP and without distorting the forward-path
measurements with reverse-path responses. The measurements can be
conducted either from $C$ to $S$ or from $S$ to $C$ to allow diagnosis
of both channel directions; we focus
on the former. Note that some APs or terminals that are not a part
of our WLAN may be nearby (e.g., in other home networks)
creating hidden terminals and/or interference, while the user has
no control over these networks.

\begin{figure}
\begin{centering}
\includegraphics[width=0.3\textwidth]{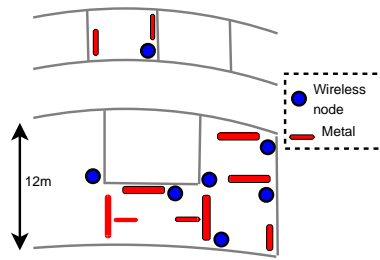}
\par\end{centering}
\caption{Testbed layout.\label{fig:testbed.}}
\end{figure}

We have conducted all experiments in this paper using 
a testbed that consists of 802.11g Soekris \texttt{net4826} nodes with
mini-PCI interfaces. The mini-PCI interfaces host either an Atheros
chipset or an Intel 2915ABG chipset, with the MadWiFi and ipw2200
drivers respectively (on the Linux 2.6.21 kernel). The MadWiFi driver
allows us to choose between four rate adaptation modules.
We disable the optional MadWiFi features referred to as
{\em fast frames} and {\em bursting} because  
they are specific to MadWiFi's \emph{Super-G} implementation and
they can interfere with the proposed rate inference process. The testbed
is housed in the College of Computing at Georgia Tech, and the
geography is shown in Figure \ref{fig:testbed.}.

%\paragraph*{{\bf Related work}}
\subsection*{Related work}
There is significant prior work in the area of WLAN monitoring and
diagnosis. However, to the extent of our knowledge, there is no earlier
attempt to diagnose WLAN problems using exclusively user-level active
probing, without any information from 802.11 devices and other layer-2 monitors. 
User-level active probing has been used to estimate {\em conflict graphs}
and hidden terminals, assuming that the involved devices cooperate
in the detection of hidden terminals  
\cite{ahmed2008online,niculescu2007interference,padhye2005estimation}.
Instead, with WLAN-probe, hidden terminals may not 
participate in the detection process (and they may be located in different WLANs).  
Passive measurements have also been used for the construction of 
conflict graphs \cite{cai2009non,giustinianomeasuring,vutukuru2008harnessing}.
Earlier systems require multiple 802.11 monitoring
devices \cite{cheng2007automating,cheng2006jigsaw,mahajan2006analyzing},
NIC-specific or driver-level support for layer-2 information 
\cite{chandra2006wifiprofiler,sheth2006mojo},
and network configuration data \cite{aggarwal2009netprints}. 
Model-based approaches use transmission observations from the NIC
to predict interference 
\cite{kashyap2007measurement,lee2007rss,qiu2007general,Reis:2006:MMD:1151659.1159921}.
Signal processing-based approaches decode PHY signals to identify
the type of interference \cite{lakshminarayanan2009rfdump}; some
commercial spectrum analyzers \cite{airmagnet,aruba} deploy such
monitoring devices at vantage points.

\section{Wireless Access Delay}\label{sec:Access-Delays}
The proposed diagnostics are based on a certain component of a probing
packet's One-Way Delay (OWD), referred to as \emph{wireless access delay} or simply
\emph{access delay}. Intuitively, this term captures the following delay 
components that a packet encounters at an 802.11 link: a) waiting for the channel
to become available, b) a (variable) backoff window before its transmission,
c) the transmission delay of potential retransmissions, and d) certain constant delays
(DIFS, SIFS, transmission of ACKs, etc).  The access
delay does \emph{not} include the potential queueing delay at the
sender due to the transmission of earlier packets, as well as the
latency for the first transmission of the packet. The access delay captures important
properties of the link layer delays which allow us to distinguish between pathologies;
further, we can estimate it with user-level measurements.

\begin{figure*}
\begin{centering}
\includegraphics[width=0.95\textwidth]{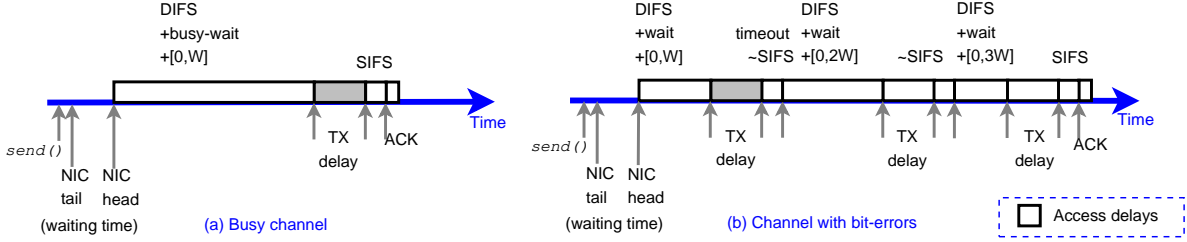}
\par\end{centering}

\caption{Timeline of an 802.11 packet transmission showing \emph{access delays}.\label{fig:Timeline-of-80211}}

\end{figure*}

Before we define the wireless access delay more precisely, let us
group the various components of the OWD $d_{i}$ of a packet $i$
from $C$ to $S$ (see Figure~\ref{fig:System-architecture.}) into
four delay components. 
We assume that the link between the AP and $S$ does not cause queueing 
delays. 
First, packet $i$ may have to \emph{wait}
at the sender NIC's transmission queue for the successful transmission
of packet $i-1$ - this is due to the FCFS nature of that queue and
it does not depend on the 802.11 protocol. If the time-distance ({}``gap'')
between the arrival of the two packets at the sender's queue is $g_{i}$,
packet $i$ will have to wait for $w_{i}$ before it is available
for transmission at the head of that queue, where: \begin{equation}
w_{i}=\max\left\{d_{i-1}-g_{i},0\right\}\label{eq:waittime}\end{equation}
We can estimate $w_{i}$ only if packet $i-1$ has
\emph{not} been lost - otherwise we cannot estimate the access delay
for packet $i$. The second delay component is the \emph{first} (and
potentially last) \emph{transmission delay} of packet $i$. In 802.11,
packets may be retransmitted several times and each transmission can
be at a different layer-2 rate in general. The ratio $s_{i}/r_{i,1}$
represents the first transmission's delay, where $s_{i}$ is the size
of the packet (including the 802.11 header and the frame-check sequence)
and $r_{i,1}$ is the layer-2 rate of the first transmission; we focus
on the estimation of $r_{i,1}$ in the next section. The third delay
component $c$ includes various \emph{constant latencies} during the
first transmission of a packet; without going into the details (which
are available in longer descriptions of the 802.11 standard), these
latencies include various DIFS/SIFS segments and the layer-2 ACK transmission
delay (which is always at the same rate). Finally, there is a \emph{variable
delay} component $\beta_{i}$. When the packet is transmitted only
once, $\beta_{i}$ consists of the waiting time ({}``busy-wait'')
for the 802.11 channel to become available as well as a random backoff
window (uniformly distributed in a certain number of time \emph{slots}).
If the packet has to be transmitted more than once, $\beta_{i}$ also
includes \emph{all the additional delays} because of subsequent retransmission
latencies, busy-wait, backoff times and constant latencies. These delay components are illustrated in Figure \ref{fig:Timeline-of-80211}.
We define the wireless access delay $a_{i}$ as 
\begin{equation}
a_{i}=c+\beta_{i}\label{eq:adelay2}\end{equation}
and so it can be estimated from the OWD as 
\begin{equation}
a_{i}=d_{i}-w_{i}-\frac{s_{i}}{r_{i,1}}\label{eq:adelay}\end{equation}
where $w_{i}$ is derived from Equation \ref{eq:waittime}.

Another way to think about the wireless access delay is as follows.
Suppose that we compare the OWD of a packet that traverses an 802.11
link with the OWD of an equal-sized packet that goes through a work-conserving
FCFS queue with constant service rate $r$ (e.g., a DSL or a switched
Ethernet port). The OWD of the latter would include the sender waiting
time $w_{i}$ and the transmission latency $s_{i}/r$. In that case
the term $a_{i}$ would only consist of the queueing delay due to
cross traffic that arrived at the link before packet $i$. In the
case of 802.11, the link is \emph{not} work-conserving (packets may
need to wait even if the channel is available), the transmission rate
can change across packets, and there may be retransmissions of the
same packet. Thus, the wireless access delay captures not only the
delays due to cross traffic, but also all the additional delays due
to the idiosyncrasies of the wireless channel and the 802.11 protocol.
A significant increase in the access delay of a packet implies either
long busy-waiting times due to cross traffic, or problematic wireless
channel conditions due to low SNR, interference etc. In the following
sections we examine the information that can be extracted from either
temporal correlations in the access delay, or from the dependencies
between access delay and packet size. It should be noted that 
the {\em access delay} can have additional applications in other
wireless network inference problems (such as available bandwidth estimation),
which we plan to investigate in future work.

%\paragraph*{{\bf Diagnosis tree and probing structure}}
\subsection*{Diagnosis tree and probing structure}
\begin{figure}
\begin{centering}
\includegraphics[width=0.4\textwidth]{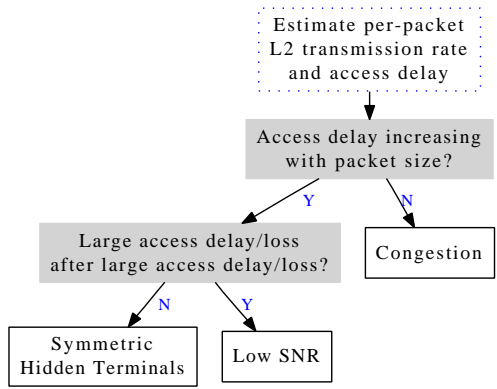}
\par\end{centering}
\caption{WLAN-probe decision tree.\label{fig:Decision-tree}}
\end{figure}

Having defined the key metric in the proposed method, we now 
present an overview of the WLAN-probe diagnosis tree that allows us
to distinguish between pathologies (see Figure \ref{fig:Decision-tree}).
We start by analyzing each packet train separately, and use a novel
dispersion-based method to infer the per-packet layer-2 transmission rate,
when possible (Section \ref{sec:Rate-Inference}). Based on the
inferred rates, we can estimate the wireless access delay for
each packet. We then examine whether the access delays increase with the packet size 
(Section \ref{sec:Low-Signal-Strength}).
When this is \emph{not} the case, the WLAN pathology is diagnosed
as congestion. 
On the other hand, when the access delays increase with the
packet size, the observed pathology is due to low SNR or hidden terminals.
We distinguish between these two pathologies based on temporal correlation
properties of packets that either encountered very
large access delays or that were lost at layer-3 
(Section~\ref{sec:Hidden-Terminals-and}).

To conduct the previous diagnosis tests, we need to probe
the WLAN channel with multiple packet trains and with packets of 
different sizes. 
Each train provides a unique ``sample'' - we need multiple
samples to make any statistical inference. 
Each train consists of several back-to-back packets of different 
sizes. The packets have to be transmitted back-to-back so that we
can use dispersion-based rate inference methods, and they have 
to be of different sizes so that we can examine the presence of
an increasing trend between access delay and size.  Specifically,
the probing phase consists of 100 back-to-back UDP packet trains.
These packet trains are sent from the WLAN-probe client $C$ to the
WLAN-probe server $S$. The packets are timestamped at $C$ and $S$
so that we can measure their \emph{relative} One-Way Delay (OWD) variations.
The two hosts do not need to have synchronized clocks, and we compensate for
clock skew during each train by subtracting the minimum OWD in that train. 
The send/receive timestamps are obtained at user-level.
There is an idle time of one second between successive packet
trains. Each train consists of 50 packets of different sizes.
About 10\% of the packets, randomly chosen, are of the minimum-possible size
(8-bytes for a sequence number and a send-timestamp, together with
the UDP/IP headers) and they are referred to as \emph{tiny-probes}
- they play a special role in transmission rate inference 
(see Section \ref{sec:Rate-Inference}). The size
of the remaining packets is uniformly selected from the set of values
$\{8+200\times k,k=1\dots7\}$ bytes.

\section{Transmission Rate Inference}\label{sec:Rate-Inference}

The computation of the wireless access delay requires the estimation
of the rate $r_{i,1}$ for the \emph{first transmission} of each probing
packet. Even though capacity estimation using packet-pair dispersion
techniques in wired networks has been studied extensively 
\cite{dovrolis2004packet,kapoor2004capprobe},
the accuracy of those methods in the wireless context has been repeatedly
questioned \cite{Portoles-Comeras:2009:ITC:1644893.1644941}. There
are three reasons that capacity estimation is much harder in the wireless
context and in 802.11 WLANs in particular. First, different packets
can be transmitted at different rates (i.e., time-varying capacity).
Second, the channel is not work-conserving, i.e., there may be idle
times even though one or more terminals have packets to send. Third,
potential layer-2 retransmissions increase the \emph{dispersion} between
packet pairs, leading to underestimation errors. On the other hand,
there are two positive factors in the problem of 802.11 transmission
rate inference. First, there are only few standardized transmission
rates, and so instead of estimating an arbitrary value we can select
one out eight possible rates. Second, most (but not all) 802.11 rate
adaptation modules show strong temporal correlations in the transmission
rate of back-to-back packets. In the following, we propose a transmission
rate inference method for 802.11 WLANs. Even though the basic idea
of the method is based on packet-pair probing, the method is novel
because it addresses the previous three challenges, exploiting these
two positive factors.

\textbf{Approach:} Recall that WLAN-probe sends many packet trains from $C$ to $S$,
and each train consists of 50 back-to-back probing packets (i.e.,
49 packet-pairs). Consider the packets $i-1$ and $i$ for a certain
train; we aim to estimate the rate $r_{i,1}$ for the first transmission
of packet $i$ given the {}``dispersion'' (or interarrival) $\Delta_{i}$
between the two packets at the receiver $S$. Of course this is possible
only when neither of these two packets is lost (at layer-3). Further,
we require that packet $i$ is \emph{not} a {}``tiny-probe''.

Let us first assume that packet $i$ was transmitted only once. In
the case of 802.11, and under the assumption of no retransmissions
for packet $i$, the dispersion can be written as: \begin{equation}
\Delta_{i}=\frac{s_{i}}{r_{i,1}}+c+\beta_{i}\end{equation}
using the notation of the previous section. To estimate $r_{i,1}$,
we first need to subtract from $\Delta_{i}$ the constant latency
term $c$ and the variable delay term $\beta_{i}$ which captures
the waiting time for the channel to become available and a uniformly
random backoff period. The sum of these two terms $c+\beta_{i}$ is
estimated using the tiny-probes; recall that their IP-layer size is
only 8 bytes and so their transmission latency is small compared to
the transmission latency for the rest of the probing packets. On the
other hand, the tiny-probes still experience the same constant latency
$c$ as larger packets, and their variable-delay $\beta$ follows
the same distribution with that of larger probing packets (because
the channel waiting time, or the backoff time, do not depend on the
size of the transmitted packet). So, considering only those packet-pairs
in which the second packet is a tiny-probe, we measure the \emph{median
dispersion} $\Delta_{\mbox{tiny}}$. This median is used as a rough
estimate of the sum $c+\beta_{i}$, when packet $i$ is \emph{not}
a tiny-probe.
\footnote{This estimate is revised in the last stage of the algorithm, after
we have obtained a first estimate for the transmission rate during
a train. We then estimate the transmission latency of each tiny-probe
and subtract it from its measured dispersion.} We then estimate the 
transmission rate $r_{i,1}$ as: \begin{equation}
r_{i,1}=\frac{s_{i}}{\Delta_{i}-\Delta_{\mbox{tiny}}}\end{equation}

If the $i$'th packet was retransmitted one or more times, the dispersion
$\Delta_{i}$ will be larger than $s_{i}/r_{i,1}+c+\beta_{i}$ and
the rate will be underestimated. A first check is to examine whether
the estimated $r_{i,1}$ is significantly smaller than the lowest
possible 802.11 transmission rate (1Mbps). In that case, we reject
the estimate $r_{i,1}$ and \emph{flag} that packet. Of course it
is possible that some remaining packets have been retransmitted, but
without being flagged at this point. We also flag all tiny-probes,
as well as any packet $i$ if packet $i-1$ was lost.

The next step is to map each remaining estimate $r_{i,1}$ to the
nearest standardized 802.11 transmission rate $\hat{r}_{i,1}$. For
instance, if $r_{i,1}$=10.5Mbps, the nearest 802.11 rate is 11Mbps.
(note that this transmission rate applies to the 802.11 frame and
so $s_{i}$ has to include the layer-2 headers).

\begin{figure}
\begin{centering}
\includegraphics[angle=270,width=0.45\textwidth]{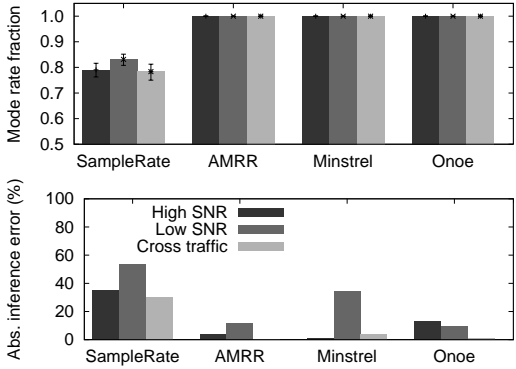}
\par\end{centering}
\caption{Rate inference: strong temporal correlations between the transmission rate of packets in the same train (top) and rate inference accuracy. Low-SNR conditions are created by separating $C$ and its AP by several meters; congestion is caused by a UDP bulk-transfer over a second network that is in-range.}
\label{fig:Rate-inference:error}
\end{figure}

We also exploit the temporal correlations
between the transmission rate of successive packets (within the same
train) to improve the existing estimates and to produce an estimate
for all flagged packets. We have experimented with the four rate adaptation
modules available in the MadWiFi driver used with the Atheros chipset
(SampleRate, AMRR, Onoe and Minstrel).
Figure~\ref{fig:Rate-inference:error} (top)
shows the fraction of probing packets in a train that were transmitted 
at the most common transmission rate during that train, 
under three different channel conditions. These
results were obtained from 100 experiments with 50-packet trains;
we also show the Wilcoxon 95\% confidence interval
in each case. Note that all rate adaptation modules exhibit strong
temporal correlations, while three of them (AMRR, Minstrel and Onoe)
seem to use a single rate for all packets during a train (each train
lasts for 5-250msec, depending on the transmission rate). 

Based on the previous strong temporal correlations, we compute
the \emph{mode} $\tilde{r}$ (most common value) of the discrete $\hat{r}_{i,1}$
estimates. If the mode includes less than a fraction (30\%) of the
measurements, we reject that packet train as \emph{too noisy}. Otherwise,
we replace every estimate $\hat{r}_{i,1}$, and the estimate for every
flagged packet, with $\tilde{r}$. If most trains show weak modes
(i.e., a mode with less than 30\% of the measurements), we abort the
diagnosis process because the underlying rate adaptation module does
not seem to exhibit strong temporal correlations between successive
packets. In our experiments, this is sometimes the case with the SampleRate
MadWiFi module. In the rest of this work, we only use 
that rate adaptation module (which is also the default in MadWiFi) 
because we want to {\em examine whether the 
proposed diagnostics work reliably even under considerable rate 
estimation errors.} 

\textbf{Evaluation:} Figure~\ref{fig:Rate-inference:error} (bottom) shows the accuracy of the
proposed rate estimation method under three quite different channel
conditions. In particular, we show the {\em average
of the absolute relative error} across all probing packets for which we
know the ground-truth transmission rate.
The {}``ground-truth'' for each packet
was obtained using an AirPcap monitor, positioned close to the sender,
that captured most (but not all) probing packets. We detect the
first transmission for each packet using the {}``Retry'' flag in
the 802.11 header. We see that the inference error is low in most cases;
the SampleRate module gives a relatively higher error.

\section{Detecting Size-dependent\\Pathologies}\label{sec:Low-Signal-Strength}
The first {}``branching point'' in the decision tree of Figure~\ref{fig:Decision-tree}
is to examine \emph{whether access delays increase with the size of
probing packets}. Recall that each probing train consists of packets
with eight distinct sizes. The pathologies in an 802.11 WLAN can be
grouped in two categories: a) pathologies that are more likely to
increase the access delay of larger packets, because of increased
waiting at the sender or increased retransmission likelihood, and
b) pathologies that increase the access delay of all packets with
the same likelihood, independent of size. We refer to the former as
\emph{ size-dependent pathologies} and the latter as \emph{size-independent}.

The first category includes a broad class of problems such as bit
errors due to noise, fading, interference, low transmission signal
strength, or hidden terminals. In the simplest (but unrealistic) case
of independent bit errors, the probability that a frame of size $s$
bits will be received with bit errors when the bit-error rate
is $p$ is $1-\left(1-p\right)^{s}$, which increases sharply with
$s$. Of course, in practice bit errors are not independent and 802.11
frame transmissions are \emph{partially} protected with FEC and rate
adaptation techniques. We expect however that when the previously
mentioned pathologies are severe enough to cause performance problems,
larger packets have a higher probability of being retransmitted, causing
an increasing trend between access delay and packet size.

The size-independent class includes pathologies that can also cause
large access delays, due to increased waiting at the sender or retransmissions,
but where the magnitude of the access delay is independent of the
packet size. The best instance in this class is WLAN congestion. It
is important, however, that the traffic that causes congestion is
generated by WLAN terminals that can {}``carrier-sense'' each other
(otherwise we have hidden-terminals). In the case of congestion, the
access delays will be larger than the case when there is no congestion
(packets have to wait more for the channel to become available) but
the access delays would not depend on the packet size.

\begin{figure}
\begin{centering}
\includegraphics[angle=270,width=0.5\textwidth]{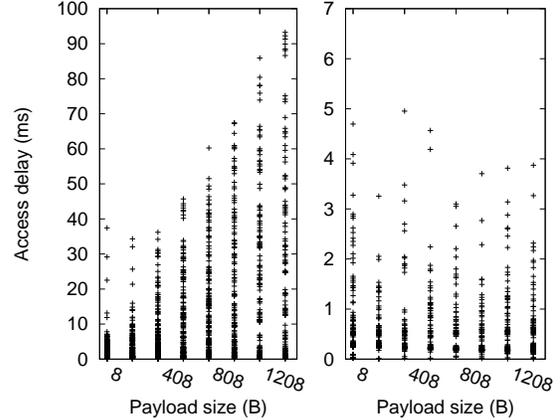}
\par\end{centering}

\caption{Low signal strength and congestion: effect of packet size (SampleRate
module).\label{fig:Low-signal-strength-CT}}

\end{figure}

\textbf{Approach:} We distinguish between the two pathology classes using statistical
trend detection in the relation between access delay and packet size.
Figure~\ref{fig:Low-signal-strength-CT}
shows the inferred access delays from experiments with 100 packet
trains. In the first experiment (left), the client $C$ and the AP
are separated by a large distance of 5-6m, so that $C$'s bulk-transfer
throughput drops to about 1Mbps. In the second experiment, we attempt
to saturate the WLAN with UDP traffic that originates from another
terminal. All terminals and APs can carrier-sense each other (we test
this based on throughput comparisons when one or more nodes are active). We use
802.11g channel-6 and SampleRate in both experiments.

%
\begin{comment}
%
\begin{figure*}
\begin{centering}
\subfloat[Low signal strength.]{\includegraphics[angle=270,width=0.45\textwidth]{figures/srate-coffeecouch-16dbm-0Mbps.pcap-ad-sz.ps}

}\subfloat[Cross traffic (10-15Mbps).]{\includegraphics[angle=270,width=0.45\textwidth]{figures/srate-16dbm-16dbm-30Mbps.pcap-ad-sz.ps}

}
\par\end{centering}

\caption{Low signal strength and congestion (SampleRate module).}

\end{figure*}

\end{comment}
{}

The access delays in the case of low signal strength increase
with the packet size, while this is not true in the case of congestion.
A more thorough analysis of these measurements reveals that \emph{not
all} access delays increase with the packet size, under low signal
strength. Instead, \emph{the increasing trend is clearly observed
among those packets that have the larger access delays for each probing
size}. This is not surprising: the packets with the larger access
delays among the set of packets of a certain size, are typically those
that are retransmitted, and the retransmission probability increases
with the packet size under size-dependent pathologies. For this reason,
instead of examining the average or the median access delay for each
packet size, we consider instead the 95-th percentile $\tilde{a}_{95p}(s)$
of the access delays for each packet size $s$.

The trend detection is performed using the nonparametric Kendall one-sided
hypothesis test \cite{hollander1999nonparametric}. The null hypothesis
is that there is no trend in the bivariate sample $\{s,\tilde{a}_{95p}(s)\}$
for $s=\{8+k\times200,k=1\dots7\}$ (bytes), while the alternate hypothesis
is that there is an increasing trend. 

\textbf{Evaluation:} For the experiments of Figure~\ref{fig:Low-signal-strength-CT}
the test strongly rejects the null hypothesis under low signal strength
with a p-value of 0 (the p-value is less than 0.01 across all MadWiFi rate
modules), while the p-value in the case of congestion is 0.81 (0.7-1.0
across all MadWiFi rate modules). We have repeated similar experiments
with all other MadWiFi rate adaptation modules and under different
signal strengths and congestion levels. The p-values in all experiments
show a clear difference between size-dependent and size-independent
pathologies, as long as the received signal strength is less than
about 8-10dBm. For higher signal strengths, the user-level throughput
is more than 5Mbps, and so it is questionable whether
there is a pathology that needs to be diagnosed in the first place.

\section{Low SNR and Hidden Terminals}\label{sec:Hidden-Terminals-and}
After the detection of a size-dependent pathology, WLAN-probe attempts
to distinguish between \emph{low-SNR conditions} and 
\emph{Symmetric Hidden Terminals (SHTs)}. The former represents a wide range of problems
(low signal strength, interference from non-802.11 devices, significant
fading, and others) - a common characteristic is that they are all
caused by \emph{exogenous factors} that affect the wireless channel
independent of the presence of traffic in the channel. SHTs represent
the case that at least two 802.11 senders (from the same or different
WLANs) can not carrier-sense each other and when they both transmit
at the same time neither sender's traffic is correctly received. SHTs
do not represent an exogenous pathology because the problem disappears
if all but one of the colliding senders backoff.
The case of asymmetric HTs (or one-node HTs), where one sender's transmissions
are corrupted while the conflicting sender's transmissions are correctly
received, is no different than the exogenous factors we consider and
WLAN-probe will diagnose them as low-SNR.

\textbf{Approach:} To distinguish between low-SNR and SHTs, we first introduce some additional
terminology of events that probing packets may see. A probing packet
may be \emph{lost at layer-3} (denoted by L3), after a number of unsuccessful
retransmissions at layer-2. A probing packet may see an \emph{outlier
delay} (OD), if its access delay is significantly higher than the
typical access delay in that probing experiment - we classify a packet
as OD if its access delay is larger than the sample median plus three
standard deviations (the sample includes all measured access delays
in that probing experiment - across all trains). 
Finally, a probing packet may see a \emph{large
delay} (LD) if its access delay is higher than the \emph{typical}
access delay in that probing experiment - we classify a packet as
LD if its access delay is higher than the 90-th percentile of the
empirical distribution of access delays (after we have excluded OD
packets). Note that the access delays of OD packets are typically
much larger than the access delays of LD packets.

\begin{figure}
\begin{centering}
\includegraphics[angle=270,width=0.45\textwidth]{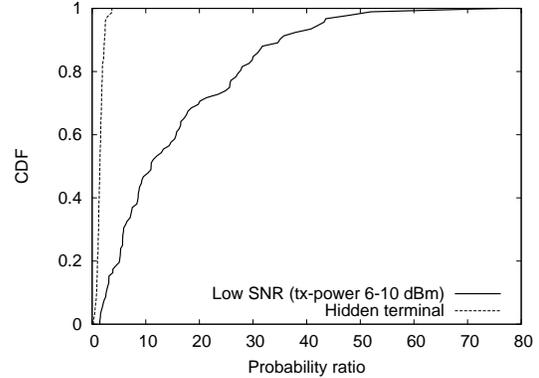} 
\par\end{centering}

\caption{Probability ratio ($p_{c}/p_{u}$) to distinguish between low-SNR and SHT conditions.\label{fig:snr-ht}}

\end{figure}

The probing and diagnosis process works as follows. The probing packets
in this WLAN-probe experiment are of the largest possible size that
will not be fragmented. The reason is that larger packets are more
likely to collide with other transmissions in the case of SHTs. We
then identify all OD or L3 packets in the probing trains of the experiment,
and estimate the unconditional probability $p_{u}$ that either event
takes place: \begin{equation}
p_{u}=\mbox{Prob}\left[\mbox{OD}\lor\mbox{L3}\right]\label{eq:pu}\end{equation}
We then focus on the \emph{successor} of an OD or L3 event, i.e.,
the probing packet that follows an OD or L3 packet. Under low-SNR
scenarios we expect that the channel conditions exhibit strong temporal
correlations, and so if a packet $i$ experiences an OD or L3 event,
its successor packet $i+1$ (denote by \emph{successor($i$)}) will
see a large delay (LD) or layer-3 loss (L3) event with high probability.

On the other hand, if packet $i$ experiences an outlier delay (OD)
or L3 event due to an SHT, the colliding senders will backoff for
a random time period and it is less likely that the successor packet
will be LD or L3. To capture the previous temporal correlations between
an L3 or OD packet and its successor, we consider the conditional
probability: \begin{equation}
p_{c}=\mbox{Prob}\left[\mbox{successor}(i):\mbox{LD}\lor\mbox{L3}\mid i:\mbox{OD}\lor\mbox{L3}\right]\label{eq:prob-ratio}\end{equation}

The detection method focuses on the ratio $p_{c}/p_{u}$ of the previous
conditional and unconditional probabilities. If there is a strong
temporal correlation between a probing packet that experiences an
OD or L3 event and its successor, this probability ratio will be much
larger than one. We expect this to be the case under low-SNR conditions.
Otherwise, under an SHTs condition, the previous temporal correlation
is much weaker and the probability ratio will be closer to one.

\textbf{Evaluation:} Figure~\ref{fig:snr-ht} shows the distribution of the probability
ratio $p_{c}/p_{u}$ for 100 low-SNR and 90 SHT experiments in our
testbed. We create low-SNR conditions by reducing the transmission
power of the WLAN-probe client $C$ to 6-10dBm; the access point is
about 3m away. We create SHTs conditions using two different networks
on 802.11g channel-6, such that the two senders can not carrier-sense
each other. When only one sender is active, the throughput in the
corresponding network is higher than 10Mbps - when both senders are
always backlogged, the throughput drops to less than 1Mbps. The probability
ratio is always less than 5 under SHTs, while it is higher than 5
in 80\% of the experiments under low-SNR conditions. A probability
ratio threshold between 3-5 should be sufficient to diagnose almost
all SHTs accurately. Under low-SNR conditions, however, we should expect
some diagnosis errors: in 10-20\% of the cases, WLAN-probe will diagnose
a low-SNR condition as SHT. We are investigating ways to further improve
the accuracy of this diagnostic process.

\section{Conclusions and future work}
We proposed a home WLAN diagnosis process that
only requires user-level active probing,
and presented some preliminary but promising
results that show the feasibility of such diagnostics.
A design consideration for our methods is \emph{usability}:
we do not require administrative privileges, any form of
support from the wireless card/driver/AP, or sensor
nodes at vantage points in the home.

We are working on several extensions of WLAN-Probe. 
First, it is possible that there is no real WLAN pathology - we are
working on a method that can distinguish between normal operation
and the previous pathologies. 
Second, some preliminary work shows that we can detect certain 
non-802.11 interference sources, such as microwave ovens.
Third, we are working on improvements in the rate inference method
and on testing these methods with additional rate adaptation mechanisms.
Finally, we will conduct a larger-scale evaluation
of the WLAN-probe diagnostic accuracy with more testbed experiments
as well as with actual home WLAN deployments.

\balance

\bibliographystyle{plain}
\bibliography{ref2}

\begin{thebibliography}{10}

\bibitem{WRAPI}
{WRAPI: API for Real-time Monitoring and Control of an 802.11 Wireless LAN}.
\newblock http://sysnet.ucsd.edu/pawn/wrapi, 2002.

\bibitem{airmagnet}
{AirMagnet WiFi Analyzer}.
\newblock http://www.airmagnet.com, 2010.

\bibitem{aruba}
{Aruba Networks: RFProtect Spectrum Analyzer}.
\newblock http://www.arubanetworks.com, 2010.

\bibitem{aggarwal2009netprints}
B.~Aggarwal, R.~Bhagwan, T.~Das, S.~Eswaran, V.N. Padmanabhan, and G.M.
  Voelker.
\newblock {NetPrints: Diagnosing home network misconfigurations using shared
  knowledge}.
\newblock In {\em USENIX NSDI}, 2009.

\bibitem{ahmed2008online}
N.~Ahmed, U.~Ismail, S.~Keshav, and K.~Papagiannaki.
\newblock {Online estimation of RF interference}.
\newblock In {\em ACM CoNEXT}, 2008.

\bibitem{cai2009non}
K.~Cai, M.~Blackstock, M.J. Feeley, and C.~Krasic.
\newblock {Non-intrusive, dynamic interference detection for 802.11 networks}.
\newblock In {\em ACM SIGCOMM IMC}, 2009.

\bibitem{chandra2006wifiprofiler}
R.~Chandra, V.N. Padmanabhan, and M.~Zhang.
\newblock {WiFiProfiler: cooperative diagnosis in wireless LANs}.
\newblock In {\em ACM Mobisys}, 2006.

\bibitem{cheng2007automating}
Y.C. Cheng, M.~Afanasyev, P.~Verkaik, P.~Benko, J.~Chiang, A.C. Snoeren,
  S.~Savage, and G.M. Voelker.
\newblock {Automating cross-layer diagnosis of enterprise wireless networks}.
\newblock {\em ACM SIGCOMM CCR}, 37(4):25--36, 2007.

\bibitem{cheng2006jigsaw}
Y.C. Cheng, J.~Bellardo, P.~Benko, A.C. Snoeren, G.M. Voelker, and S.~Savage.
\newblock {Jigsaw: Solving the puzzle of enterprise 802.11 analysis}.
\newblock {\em ACM SIGCOMM CCR}, 36(4):39--50, 2006.

\bibitem{dovrolis2004packet}
C.~Dovrolis, P.~Ramanathan, and D.~Moore.
\newblock {Packet-dispersion techniques and a capacity-estimation methodology}.
\newblock {\em Networking, IEEE/ACM Transactions on}, 12(6):963--977, 2004.

\bibitem{giustinianomeasuring}
D.~Giustiniano, D.~Malone, D.J. Leith, and K.~Papagiannaki.
\newblock {Measuring transmission opportunities in 802.11 links}.
\newblock {\em IEEE/ACM ToN}, (99):1, 2010.

\bibitem{hollander1999nonparametric}
M.~Hollander and D.A. Wolfe.
\newblock {Nonparametric Statistical Methods}.
\newblock 1999.

\bibitem{kapoor2004capprobe}
R.~Kapoor, L.J. Chen, L.~Lao, M.~Gerla, and M.Y. Sanadidi.
\newblock {CapProbe: a simple and accurate capacity estimation technique}.
\newblock In {\em ACM SIGCOMM}, 2004.

\bibitem{kashyap2007measurement}
A.~Kashyap, S.~Ganguly, and S.R. Das.
\newblock {A measurement-based approach to modeling link capacity in
  802.11-based wireless networks}.
\newblock In {\em ACM MOBICOM}, 2007.

\bibitem{lakshminarayanan2009rfdump}
K.~Lakshminarayanan, S.~Sapra, S.~Seshan, and P.~Steenkiste.
\newblock {RFDump: an architecture for monitoring the wireless ether}.
\newblock In {\em ACM CoNEXT}, 2009.

\bibitem{lee2007rss}
J.~Lee, S.J. Lee, W.~Kim, D.~Jo, T.~Kwon, and Y.~Choi.
\newblock {RSS-based carrier sensing and interference estimation in 802.11
  wireless networks}.
\newblock In {\em IEEE SECON}, 2007.

\bibitem{mahajan2006analyzing}
R.~Mahajan, M.~Rodrig, D.~Wetherall, and J.~Zahorjan.
\newblock {Analyzing the MAC-level behavior of wireless networks in the wild}.
\newblock {\em ACM SIGCOMM CCR}, 36(4):75--86, 2006.

\bibitem{niculescu2007interference}
D.~Niculescu.
\newblock {Interference map for 802.11 networks}.
\newblock In {\em ACM SIGCOMM IMC}, 2007.

\bibitem{padhye2005estimation}
J.~Padhye, S.~Agarwal, V.N. Padmanabhan, L.~Qiu, A.~Rao, and B.~Zill.
\newblock {Estimation of link interference in static multi-hop wireless
  networks}.
\newblock In {\em ACM SIGCOMM IMC}, 2005.

\bibitem{Portoles-Comeras:2009:ITC:1644893.1644941}
Marc Portoles-Comeras, Albert Cabellos-Aparicio, Josep Mangues-Bafalluy, Albert
  Banchs, and Jordi Domingo-Pascual.
\newblock Impact of transient csma/ca access delays on active bandwidth
  measurements.
\newblock In {\em ACM SIGCOMM IMC}, 2009.

\bibitem{qiu2007general}
L.~Qiu, Y.~Zhang, F.~Wang, M.K. Han, and R.~Mahajan.
\newblock {A general model of wireless interference}.
\newblock In {\em ACM MOBICOM}, 2007.

\bibitem{Reis:2006:MMD:1151659.1159921}
Charles Reis, Ratul Mahajan, Maya Rodrig, David Wetherall, and John Zahorjan.
\newblock Measurement-based models of delivery and interference in static
  wireless networks.
\newblock {\em ACM SIGCOMM}, 2006.

\bibitem{sheth2006mojo}
A.~Sheth, C.~Doerr, D.~Grunwald, R.~Han, and D.~Sicker.
\newblock {MOJO: A distributed physical layer anomaly detection system for
  802.11 WLANs}.
\newblock In {\em ACM Mobisys}, 2006.

\bibitem{vutukuru2008harnessing}
M.~Vutukuru, K.~Jamieson, and H.~Balakrishnan.
\newblock {Harnessing exposed terminals in wireless networks}.
\newblock In {\em USENIX NSDI}, 2008.

\end{thebibliography}

\end{document}